\RequirePackage{fix-cm}
\documentclass[twocolumn]{svjour3}          % twocolumn
\smartqed  % flush right qed marks, e.g. at end of proof
\usepackage{graphicx}
\usepackage{epstopdf}
\usepackage{amsmath}
\usepackage{color}
\usepackage[english]{babel}
\usepackage{cite}
\usepackage{hyperref}

\usepackage{mathptmx}      % use Times fonts if available on your TeX system
\usepackage{latexsym}

\journalname{Applied Physics A}
\begin{document}

\title{Rippled area formed by surface plasmon polaritons upon femtosecond laser double-pulse irradiation of silicon: the role of carrier generation and relaxation processes}



\titlerunning{Rippled area on silicon upon femtosecond laser double-pulse irradiation}        % if too long for running head

\author{Thibault J.-Y. Derrien         \and
        J\"org Kr\"uger \and
	Tatiana E. Itina \and
	\\Sandra H\"ohm \and
	Arkadi Rosenfeld \and
	J\"orn Bonse
}


\institute{Thibault J.-Y. Derrien, J\"org Kr\"uger,  J\"orn Bonse\at
              BAM Bundesanstalt f\"ur Materialforschung und -pr\"ufung,
		Unter den Eichen 87, D-12205 Berlin, Germany  \\
              Tel.: +49-(0)30-8104-3562\\
              Fax: +49-(0)30-8104-1827\\
              \email{thibault.derrien@gmail.com}           %  \\
	\and
           Tatiana E. Itina  \at
              Laboratoire Hubert Curien (LabHC), UMR CNRS 5516 - Universit\'e Jean Monnet
		Bat. F, 18 rue du Professeur Benoit Lauras, 42000 Saint-Etienne, France
	\and
           Sandra H\"ohm, Arkadi Rosenfeld \at
              Max-Born-Institut f\"ur Nichtlineare Optik und Kurzzeitspektroskopie (MBI), 
Max-Born-Stra\ss e 2A, D-12489 Berlin, Germany 
}

\date{Received: date / Accepted: date}

\maketitle

\begin{abstract}
The formation of laser-induced periodic surface structures (LIPSS, ripples) upon
irradiation of silicon with multiple irradiation sequences consisting of
femtosecond laser pulse pairs (pulse duration 150 fs, central wavelength 800 nm)
is studied numerically using a rate equation system along with a two-temperature
model accounting for one- and two-photon absorption and subsequent carrier
diffusion and Auger recombination processes. The temporal delay between the
individual equal-energy fs-laser pulses was varied between $0$ and $\sim 4$ ps for
quantification of the transient carrier densities in the conduction band of the
laser-excited silicon. The results of the numerical analysis reveal the
importance of carrier generation and relaxation processes in fs-LIPSS formation
on silicon and quantitatively explain the two time constants of the delay dependent
decrease of the Low-Spatial-Frequency LIPSS (LSFL) area observed experimentally. The role of carrier generation, diffusion and recombination are quantified individually. \\DOI: 10.1007/s00339-013-8205-2. The final publication is available at \url{http://link.springer.com}. 
\end{abstract}

\section{Introduction}

The irradiation of solids with multiple linear polarized femtosecond laser pulses at fluences close
to the damage threshold leads to the formation of laser-induced periodic surface structures (LIPSS) on the surface of almost all materials \cite{Borowiec2003, Huang2009a, Chakravarty2011,
Bonse2012}. For strong absorbing materials such as metals or
semiconductors, in most cases low-spatial-frequency LIPSS (LSFL) are observed with a period
$\Lambda_{LSFL}$ close to the irradiation wavelength $\lambda$ \cite{Borowiec2003, Huang2009a,
Bonse2012}. These LSFL are generated by interference of the incident laser beam with a surface
electromagnetic wave (SEW) generated at a rough surface \cite{Sipe1983, Bonch-Bruevich1992}.

On silicon, predominantly LSFL were observed after low repetition rate ($\le 1$ kHz) Ti:Sapphire
femtosecond laser pulse irradiations in air environment \cite{Bonse2002, Costache2004,
Guillermin2007, Bonse2009, Bonse2010}. Their orientation is perpendicular to the laser beam
polarization and the periods typically range between $\sim 0.6 \lambda$ and $\lambda$\textcolor{black}{, depending on the degree of material excitation \cite{Bonse2009, Derrien2013}, and the number of laser pulses per spot \cite{Bonse2010}}. Several authors
have suggested that these structures are caused by excitation of surface plasmon polaritons (SPP) at the air - silicon interface when the material turns from a semiconducting
into a metallic state \cite{Huang2009a, Bonse2009, Martsinovskii2008}. The interference between 
 the electromagnetic field of the SPP and the incident laser pulse leads to a spatially modulated deposition of optical energy to the electronic system of the material. After coupling to the lattice system \cite{Derrien2010} and subsequent ablation processes, this results in a periodically corrugated surface topography \cite{Barberoglou2013}. 

The SPP-hypothesis has led to recent experiments investigating the impact of a temporally tailored energy distribution to the silicon surface by double-pulse irradiation \cite{Hoehm2013, Hoehm2013a, Barberoglou2013}.  In this material, the LSFL spatial period does not significantly depend on the double-pulse delay $\Delta t$
\cite{Hoehm2013a, Barberoglou2013}, while the LSFL rippled area strongly decreases with 
delays up to several ps \cite{Hoehm2013}. Two characteristic
exponential decay times of $\sim 0.15$ ps and $\sim 11$ ps were found.

In a numerical study, we have demonstrated that the SPP active area caused by a spatially Gaussian beam profile quantitatively explains the LSFL-covered
(rippled) area as a function of the double-fs-pulse delay \cite{Derrien2013a}. In this work, we extend the latter study and detail the contributions of the individual carrier generation and relaxation processes i.e.,
one- and two-photon absorption, carrier collisions and diffusion, and Auger recombination. 

\section{Theoretical model}

For SPP excitation, the silicon has to turn from a semiconducting to a metallic state upon fs-laser excitation, the following criterion has to be fulfilled \cite{Raether1986}
\begin{equation}
\Re e\left[\varepsilon_{Si}^{*}\right]<-1.\label{eq:CritereSPP}
\end{equation}
Here, $\varepsilon_{Si}^{*}$ represents the dielectric function of the laser-excited silicon which can be described (as a function of laser-induced carrier density $N_e$) by a Drude model \cite{Sokolowski-Tinten2000}
\begin{equation}
\varepsilon_{Si}^{*}(N_{e})=\varepsilon_{Si}-\frac{\omega_{p}^{2}(N_{e})}{\omega^{2}\left(1+i\frac{\nu}{\omega}\right)},\label{eq:fulldielectricfunction}
\end{equation} where $\omega_p=\sqrt{\frac{N_e e^2}{m_e^* \varepsilon_0}}$ represents the plasma frequency and $\omega$ the laser angular frequency [$e$: electron charge, $\varepsilon_0$: dielectric
permittivity of the vacuum]. 

\begin{table*}[t]
\begin{centering}
\begin{tabular}{cc|cc|c}
 & \multicolumn{1}{c}{} &  & \multicolumn{1}{c}{} & \tabularnewline
\hline 
\hline 
{\small Coefficient}  & {\small Symbol}  & {\small Value}  & {\small Unit}  & {\small Reference}\tabularnewline
\hline 
{\small Band gap energy}  & {\small $E_{g}$}  & {\small $1.12$}  & {\small eV}  & {\small \cite{Bauerle2000}}\tabularnewline
{\small Dielectric constant of crystalline silicon}  & {\small $\varepsilon_{Si}$}  & {\small $13.64+0.048i$}  & {\small $\mathrm{-}$}  & {\small \cite{Palik1985}}\tabularnewline
{\small One-photon absorption coefficient}  & {\small $\sigma_{1}$}  & {\small $1.021\times10^{5}$}  & {\small $\mathrm{m^{-1}}$}  & {\small \cite{Palik1985}}\tabularnewline
{\small Two-photon absorption coefficient}  & {\small $\sigma_{2}$}  & {\small $6.8\times10^{-11}$}  & {\small $\mathrm{m/W}$}  & {\small \cite{Sabbah2002}}\tabularnewline
\hline 
{\small Carrier collision time}  & {\small $\nu^{-1}$}  & {\small $1.1\times10^{-15}$}  & {\small s}  & {\small \cite{Sokolowski-Tinten2000}}\tabularnewline
{\small Effective optical mass}  & {\small $m_{e}^{*}$}  & {\small $1.64\times10^{-31}$}  & {\small kg}  & {\small \cite{Sokolowski-Tinten2000}}\tabularnewline
\hline 
{\small Auger recombination rate}  & {\small $C_{AR}$}  & {\small $3.8\times10^{-43}$}  & {\small $\mathrm{m^{6}/s}$}  & {\small \cite{Driel1987}}\tabularnewline
{\small Minimum Auger recombination time}  & {\small $\tau_{AR}$}  & {\small $6.0\times10^{-12}$}  & {\small s}  & {\small \cite{Yoffa1980}}\tabularnewline
\hline 
{\small Minimum electron-phonon coupling time}  & {\small $\tau_{\gamma}^{0}$}  & {\small $240\times10^{-15}$}  & {\small s}  & {\small \cite{Sjodin1998}}\tabularnewline
{\small Threshold density for electron-phonon coupling}  & {\small $N_{th}$}  & {\small $6.02\times10^{26}$}  & {\small m$^{-3}$}  & {\small \cite{Sjodin1998}}\tabularnewline
\hline 
\hline 
 & \multicolumn{1}{c}{} &  & \multicolumn{1}{c}{} & \tabularnewline
\end{tabular}
\par\end{centering}

\caption{\label{tab:DensityOfFluence}Material parameters used in the numerical
simulations of femtosecond laser-irradiated silicon (wavelength $\lambda=800$
nm, pulse duration $\tau=150$ fs).}
\end{table*}

The temporal change of the carrier density in the conduction band $\frac{\partial N_e}{\partial t}$ is described by a nonlinear partial
differential equation (Eq. \ref{eq:fcrDensity}) considering carrier generation, carrier diffusion
and Auger recombination. The carrier diffusion is driven by its temperature $T_e$, which
couples via electron-phonon (e-ph) interaction with the silicon lattice temperature $T_{Si}$ and is
described by a two-temperature model (Eqs. \ref{eq:TTMsi}, \ref{eq:TTMe}).

\begin{alignat}{1}
\frac{\partial N_{e}}{\partial t} & = \overbrace{\boldsymbol{\nabla} \left(k_{B}T_{e}\mu_{e}\boldsymbol{\nabla}N_{e}\right)}^{\text{carrier diffusion}}+\overbrace{G_{e}}^{\text{generation}} - \overbrace{R_{e}}^{\text{recombination}}\label{eq:fcrDensity}\\
C_{Si}\frac{\partial T_{Si}}{\partial t} & =\boldsymbol{\nabla}\left(\kappa_{Si}\boldsymbol{\nabla}T_{Si}\right)+\gamma\left(T_{e}-T_{Si}\right) \label{eq:TTMsi} \\
C_{e}\frac{\partial T_{e}}{\partial t} & =\underbrace{\boldsymbol{\nabla}\left(\kappa_{e}\boldsymbol{\nabla}T_{e}\right)}_{\text{heat diffusion}} - \underbrace{\gamma\left(T_{e}-T_{Si}\right)}_{\text{e-ph coupling}}+\underbrace{Q_{e}}_{\text{heat source}} \label{eq:TTMe} 
\end{alignat}

The carrier generation rate is given by
$G_{e}=\frac{\sigma_{1}I}{\hbar\omega}+\frac{\sigma_{2}I^{2}}{2\hbar\omega}$, the Auger
recombination rate by $R_{e}=\frac{N_{e}}{\tau_{AR}+{\left(C_{AR}N_{e}^{2}\right)^{-1}}}$ and the
heat source term is
$Q_{e}= \allowbreak (\hbar\omega-E_{g})\frac{\sigma_{1}I}{\hbar\omega} + \allowbreak (2\hbar\omega-E_{g})\frac{\sigma_{2}I^{2}}{2\hbar\omega}  \allowbreak + E_{g} R_{e} \allowbreak - \frac{3}{2} k_{B} T_{e}\frac{\partial N_{e}}{\partial t}$. The carrier mobility $\mu_e=e / (m_e^* \nu)$ depends on the collision frequency $\nu$. The relevant parameters are compiled in Tab. \ref{tab:DensityOfFluence}. 

The intensity $I(t,z)$ in the sample is calculated by numerically solving the equation $\frac{\partial I}{\partial z}=-(\sigma_{1}I+\sigma_{2}I^{2}). \label{eq:Beer-Lambert-1}$
At the sample surface, it is calculated via $I(t,z=0)=\left[1-R \right]I_{0}(t)$, where $R$ is the transient surface reflectivity at $\lambda = 800$ nm \cite{Bulgakova2010}. For a temporally Gaussian double-pulse, the incident laser intensity $I_0$ is given by 
\begin{equation}
I_{0}(t)=\frac{F_{0}}{\tau} \sqrt{\frac{4 \ln 2}{\pi}}\left[e^{-\frac{1}{2}\left(\frac{t}{\sigma_{\tau}}\right)^{2}}+e^{-\frac{1}{2}\left(\frac{t-\Delta t}{\sigma_{\tau}}\right)^{2}}\right], \label{eq:DoublePulse}
\end{equation}
where $F_0$ denotes the peak fluence of each individual pulse of the 
sequence. For more details on the model, refer to Refs. \cite{Derrien2013a, Bulgakova2005}. 

\begin{figure}
\begin{centering}
\includegraphics[width=9cm]{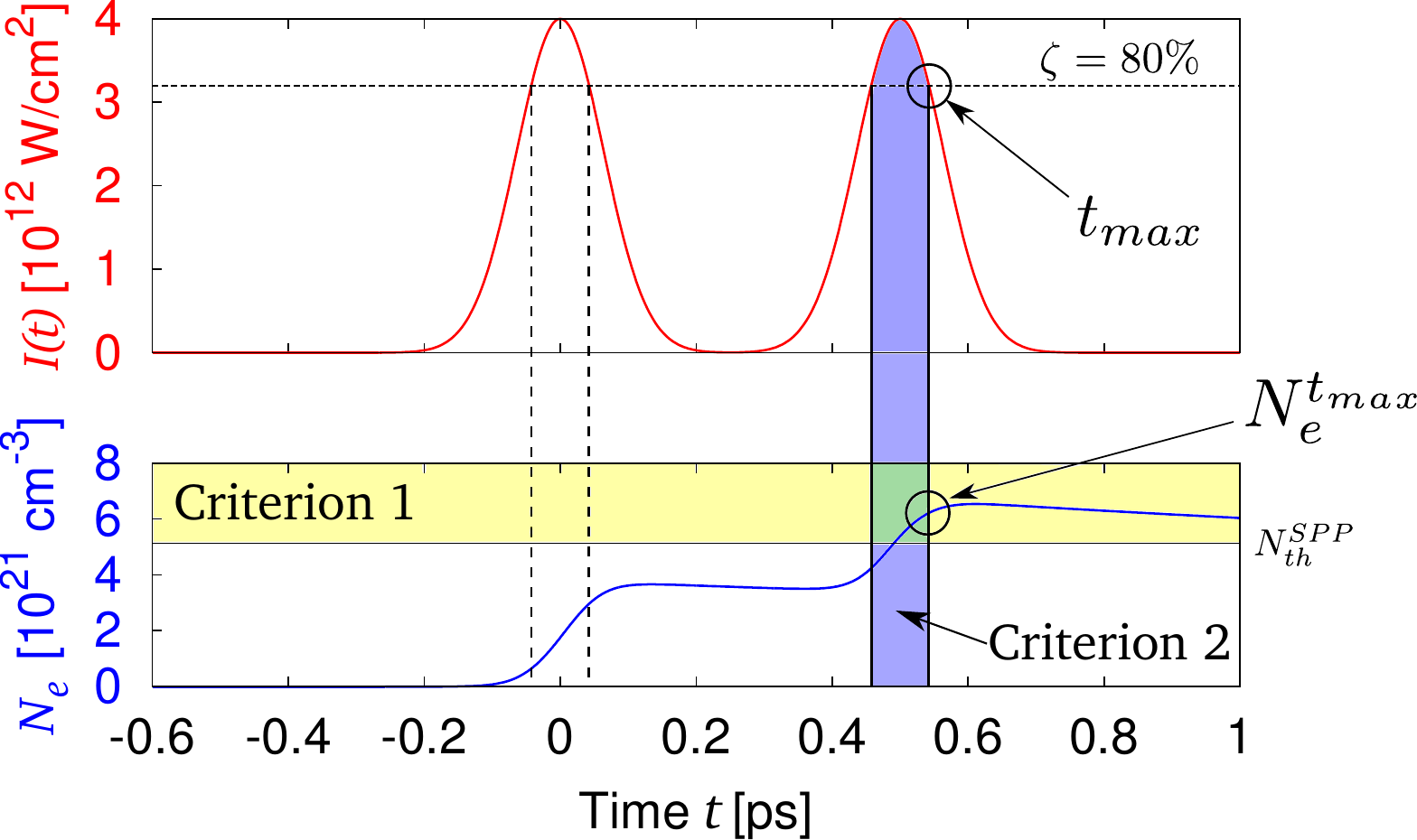}
\par\end{centering}

\caption{Illustration of the two SPP excitation criteria (as discussed in the text). The upper part shows the intensity profile of a double-pulse (Eq. \ref{eq:DoublePulse}) with $\Delta t=0.5$ ps, $\tau=150$ fs and $F_0=0.3$ J/cm$^2$. The lower part represents a numerical calculation of the corresponding carrier density.  The colored regions indicate the regimes where the criteria are satisfied individually. $t_{max}$ defines the last moment when both criteria are fulfilled simultaneously. \label{fig:TimeInterferenceCrit}}
\end{figure}

Two criteria must be satisfied to allow the excitation of Surface Plasmon Polaritons (SPPs). 
{\it Criterion 1} is defined by the condition that carrier density must exceed a threshold density $N_{th}^{SPP}$ defined by Eq.
(\ref{eq:CritereSPP}). Combining Eqs. (\ref{eq:CritereSPP}) and (\ref{eq:fulldielectricfunction}), this threshold density can be rewritten as \cite{Martsinovskii2008}
\begin{equation}
N^{SPP}_{th}=\frac{m^{*}_e\varepsilon_{0}\left(\Re e\left[ \varepsilon_{Si}\right]+1\right)}{e^{2}}\left(\omega^{2}+\nu^{2}\right). \label{eq:Nth}
\end{equation}
{\it Criterion 2} is based on the idea that temporal interference between the incident laser wave and the SPP is required (which occurs lastly during the second part of the double-pulse). This defines the last instant $t_{max}$ where the intensity of the second pulse drops to a fraction $\zeta$ of its maximum value [$I_0 (t_{max}) = \zeta \times \max \left\{ I_0 (t) \right\}=\zeta \times F_0 / \tau \times  \sqrt{4 \ln 2 / \pi} $]. 
Both criteria are illustrated in Fig. \ref{fig:TimeInterferenceCrit} where the intensity distribution of the double-pulse sequence is shown in the upper part, and the corresponding carrier dynamics is presented in the lower part. $\zeta$ is exemplified for a value of $80 \%$, and the corresponding $t_{\text{max}}$ is indicated by  the two circles. 

\begin{figure}[b]
\begin{centering}
\includegraphics[width=7cm]{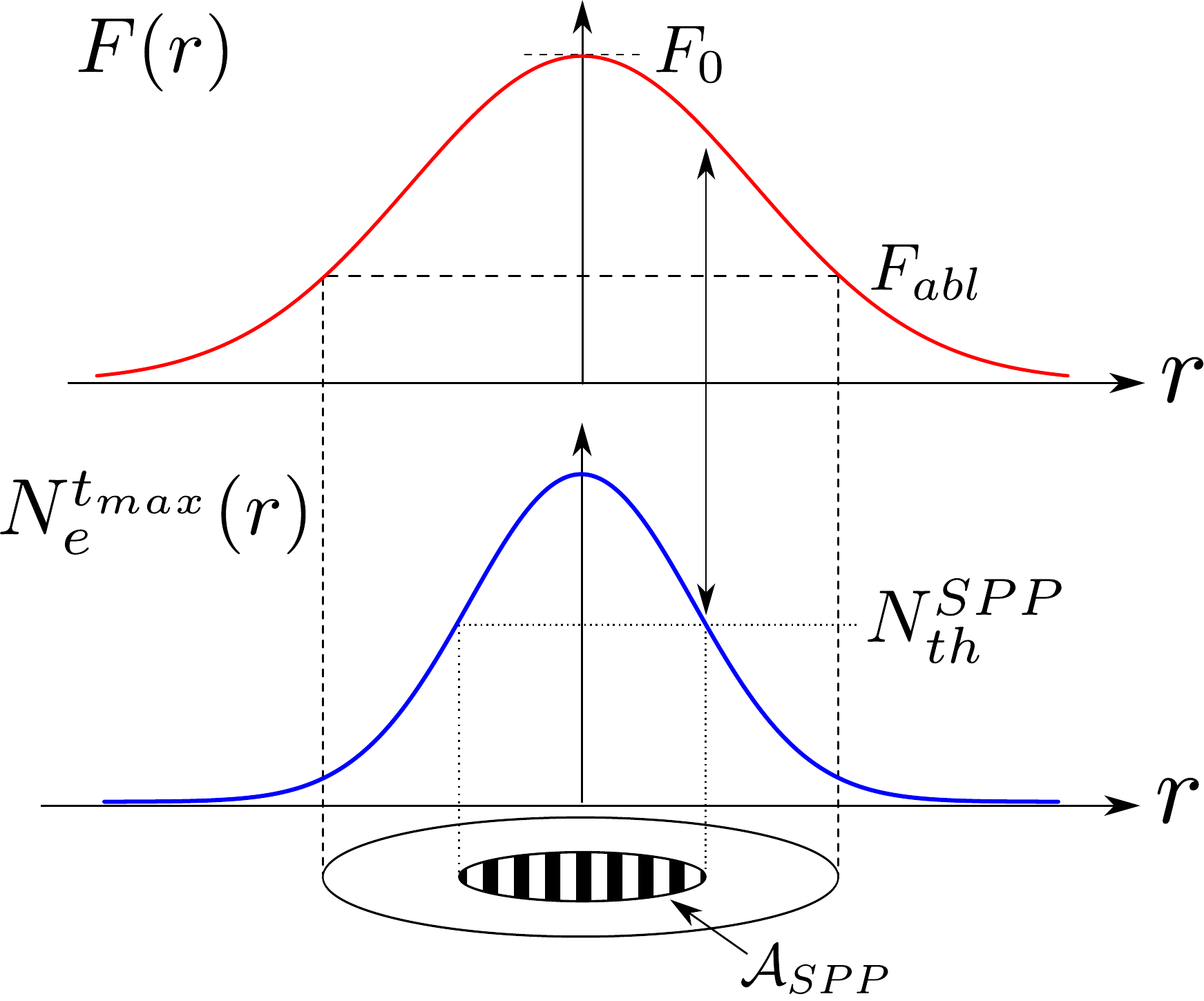}
\par\end{centering}

\caption{Scheme of the evaluation of the SPP active area $\mathcal{A}_{SPP}$ considering the spatially Gaussian beam profile (upper part), and the corresponding radial carrier density distribution (lower part). Each radial position of the Gaussian beam can be associated with a carrier density $N_e^{t_{max}}$ which respects the temporal interference criterion. The threshold density for SPP excitation $N_{th}^{SPP}$ then defines $\mathcal{A}_{SPP}$. \label{fig:AreaCalculation}}
\end{figure}

In order to consider the spatially Gaussian beam profile used in the experiments \cite{Hoehm2013}, the corresponding
radial distribution of the carrier density has to be calculated. For that, the Eqs.
(\ref{eq:fcrDensity}-\ref{eq:TTMe}) have been simultaneously solved numerically for different
fluences and delays. The radial position $r$ of the Gaussian beam is linked to the local laser fluence $F$
by the relation $F(r)=F_0 \exp [-2\left( r/w_0 \right)^2]$. Thus, the radial position $r$ can be
associated with $N_e^{t_{max}} (t,\allowbreak \Delta t, F_0)$. Applying now the threshold criterion (Eq.
\ref{eq:Nth}) allows to quantify the SPP active area $\mathcal{A}_{SPP}$, as illustrated in Fig.
\ref{fig:AreaCalculation}.
By systematically varying $\Delta t$ for the given experimental conditions, a comparison between the SPP active area $\mathcal{A}_{SPP}$ and the LSFL-rippled area $\mathcal{A}_{LSFL}$ can be performed as discussed in the following section. 

\section{Results}

In order to quantify the impact of the optical absorption, Auger recombination and carrier diffusion, all of these processes have been studied individually. For each case, starting from the optimum agreement demonstrated in Ref. \cite{Derrien2013a}, the corresponding process parameter (two-photon absorption: $\sigma_2$, Auger recombination: $R_e$, carrier diffusion: $\mu_e$) has been varied while keeping the others unchanged. 

\begin{figure}
\begin{centering}
\includegraphics[width=8cm]{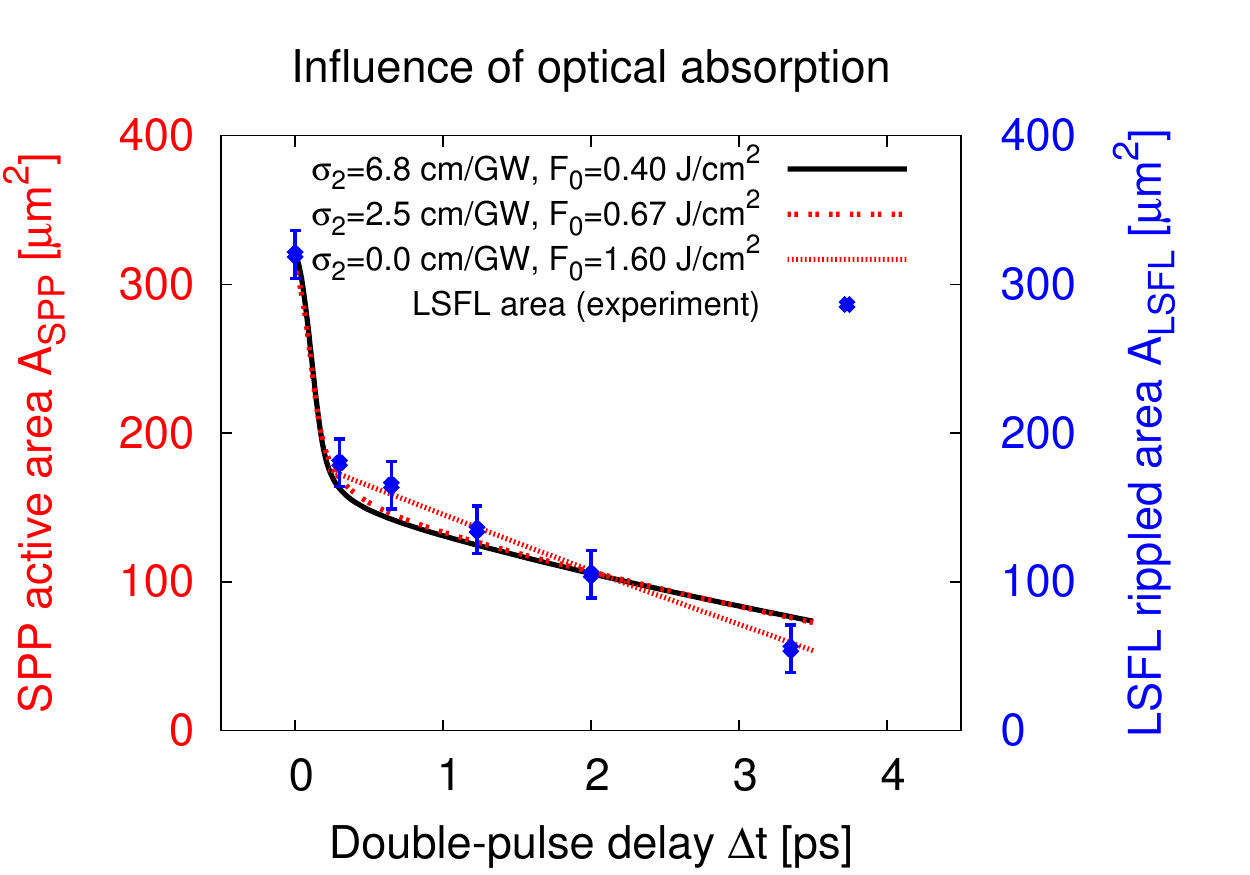}
\par\end{centering}

\caption{SPP active area as a function of double-pulse delay for different values of two-photon absorption coefficient $\sigma_2$. $F_0$ has been adjusted for an optimum agreement with experiments. In all cases, $\zeta$ has been fixed to $99\%$.\label{fig:SPPareatwophoton}}
\end{figure}

Fig. \ref{fig:SPPareatwophoton} shows the
SPP active area $\mathcal{A}_{SPP}$ as a function of the double-pulse delay $\Delta t$ up to $3.5$
ps for three different values of two-photon absorption coefficient $\sigma_2=0$, $2.5$ and $6.8$
cm/GW. The results of the numerical calculations are shown as lines, while the LSFL rippled area is
added as blue data points for comparison. The black solid line represents the optimum agreement between
$\mathcal{A}_{SPP}$ and $\mathcal{A}_{LSFL}$. For
$\sigma_2=0$ (i.e., one-photon absorption only), the results do not reproduce the rapid decay of LSFL-rippled area at short pulse
delays. Moreover, to obtain an optimum agreement with $\mathcal{A}_{LSFL}$ at longer delays, the peak
fluence had to be set to $F_0=1.6$ J/cm$^2$, which is 10 times higher than the experimental value
$F_0^{exp}=0.15$ J/cm$^2$ \cite{Hoehm2013}. In order to quantify the importance of the two-photon
absorption, $\sigma_2$ has been varied between the two most prominent values found in the
literature, i.e., $\sigma_2=2.5$ cm/GW \cite{Bristow2007} and $\sigma_2=6.8$ cm/GW \cite{Sabbah2002}.
However, no signficant differences can be observed when the peak fluence is adjusted individually
for an optimum agreement ($F_0=0.67$ J/cm$^2$ for $\sigma_2=2.5$ cm/GW and $F_0=0.40$ J/cm$^2$ for
$\sigma_2=6.8$ cm/GW). The choice of the latter is more reasonable here as its corresponding fluence is closer to
the experimental value. These results demonstrate that the two-photon absorption is essential to
explain the rapid decay of $\mathcal{A}_{SPP}$. 

\begin{figure}[b]
\begin{centering}
\includegraphics[width=8cm]{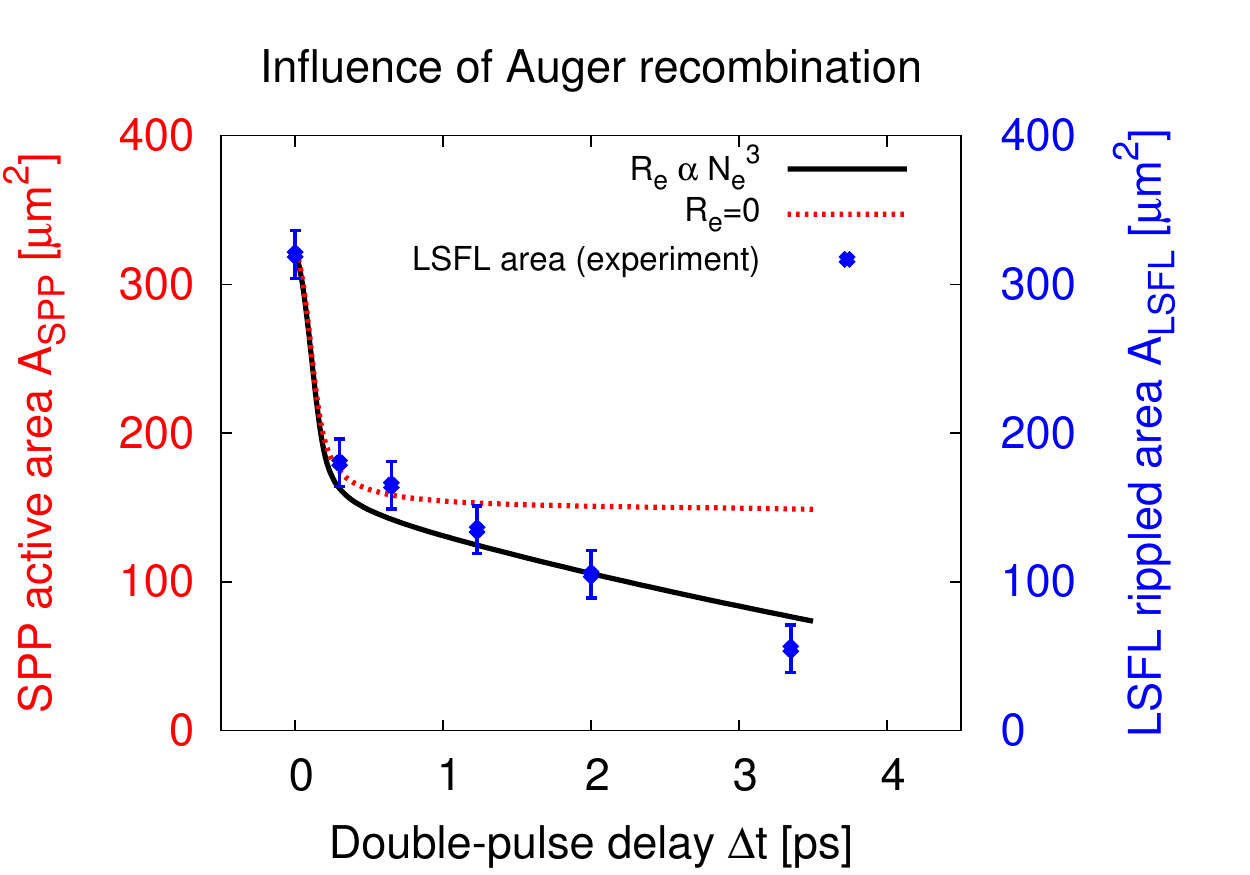}
\par\end{centering}

\caption{SPP active area as a function of double-pulse delay for Auger recombination included and excluded ($R_e=0$).  In both cases, $\zeta$ has been fixed to $99\%$ and $F_0=0.40$ J/cm$^2$.\label{fig:Auger}}
\end{figure}

In order to quantify the impact of Auger recombination, the Auger recombination rate has been set to
$R_e=0$. Fig. \ref{fig:Auger} shows the SPP active area $\mathcal{A}_{SPP}$ as a function of the
double-pulse delay $\Delta t$ with (solid line) and without (dotted line) Auger recombination. For $R_e=0$ and temporally non-overlapping double-pulses, the SPP active area remains almost constant, indicating that diffusion cannot significantly reduce the laser-induced carrier density on a timescale of a few ps. The case with Auger recombination included ($R_e \propto N_e^3$) clearly demonstrates the major contribution of this effect to the SPP active area. 

\begin{figure}
\begin{centering}
\includegraphics[width=8cm]{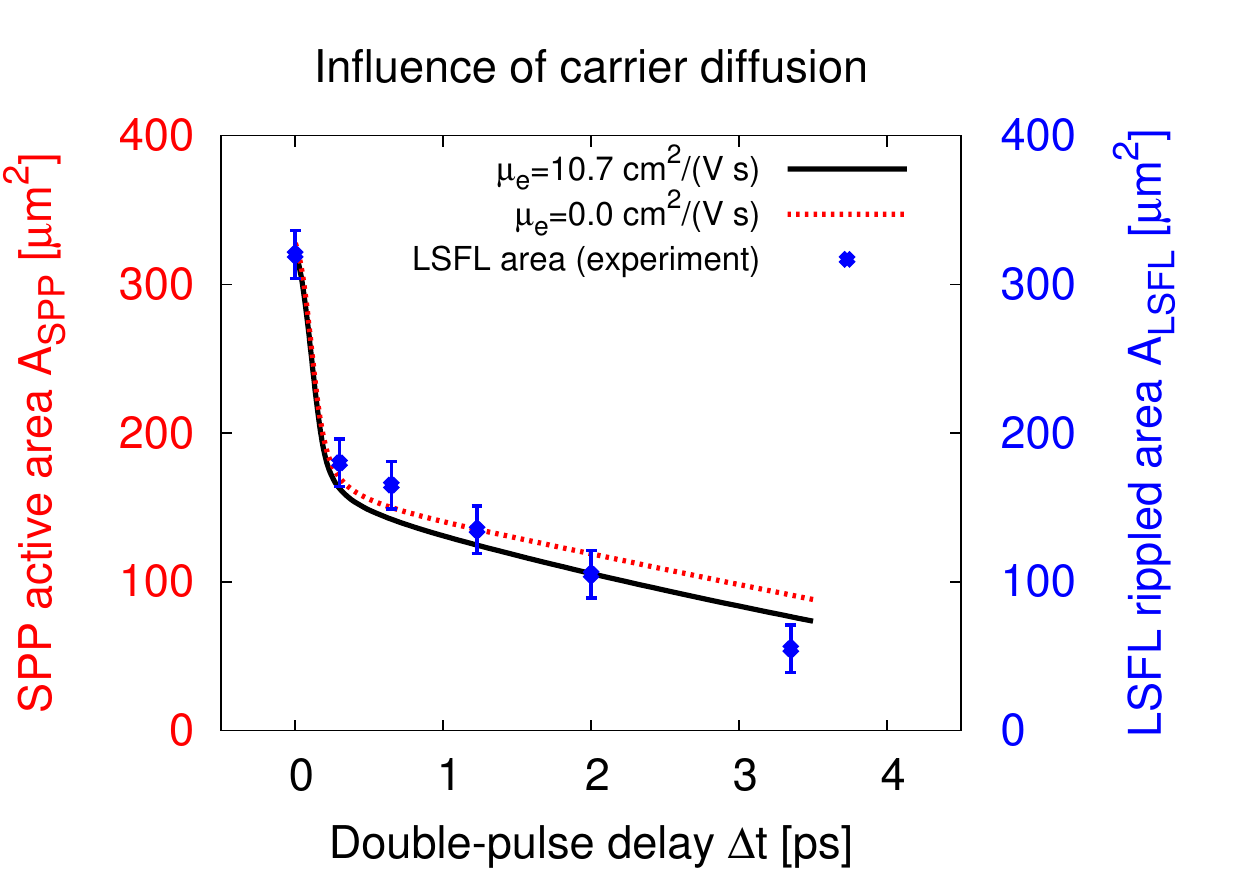}
\par\end{centering}

\caption{SPP active area as a function of double-pulse delay for two different carrier mobilies $\mu_e$.  In all cases, $\zeta$ has been fixed to $99\%$ and $F_0=0.40$ J/cm$^2$. \label{fig:AreaCarrier}}
\end{figure}

In order to study the effect of carrier diffusion, the carrier mobility $\mu_e$ was varied. Fig.
\ref{fig:AreaCarrier} shows the SPP active area $\mathcal{A}_{SPP}$ as a function of the
double-pulse delay $\Delta t$ for two different carrier mobilities $\mu_e$. The best agreement (black
curve) with $\mathcal{A}_{LSFL}$ has been obtained by setting $\mu_e=e/(m_e^* \nu) \allowbreak = 10.7$ cm$^2/(V s)$
\cite{Derrien2013a}, resulting from the carrier collision time $\nu^{-1}=1.1$ fs reported in Ref. \cite{Sokolowski-Tinten2000}. The dotted curve has been calculated by setting $\mu_e=0$. This comparison demonstrates the limited influence of carrier diffusion on SPP active area on the timescale up to a few ps. 

\section{Conclusion}

The carrier dynamics at the surface of silicon upon femtosecond double-laser pulse irradiation
has been numerically investigated as a function of double-pulse delay and laser peak fluence,
considering different processes of carrier generation and relaxation. Based on that and on two optical criteria, the SPP active
area was quantified. The comparison to experimental results of the LSFL rippled area 
confirms the SPP-based mechanism of LSFL formation. It was quantitatively demonstrated that the
two-photon absorption is responsible for the fast decay in the sub-picosecond delay range of the SPP active area, while Auger recombination accounts for the slower area decay at delays up to several
picoseconds. Diffusion plays a minor role only. 

\begin{acknowledgements}
T.J.-Y.D. acknowledges a postdoctoral fellowship awarded by the Adolf-Martens-Fond e.V. This work was supported by the German Science Foundation (DFG) under Grant Nos. RO 2074/7-2 and KR 3638/1-2. 

\end{acknowledgements}

% \bibliographystyle{spmpsci}      % mathematics and physical sciences
\bibliographystyle{spphys}       % APS-like style for physics

% \bibliography{/media/PHD/Travail/These/Articles/bibliographie_lue}

\end{document}